\documentclass{aa}  
\usepackage{graphicx}
\usepackage{txfonts}
\usepackage[colorlinks=true,
            linkcolor=blue,
            urlcolor=blue,
                citecolor=blue]{hyperref}
\begin{document} 
  
   \title{The CMB Cold Spot as predicted by foregrounds around nearby galaxies}

   
   \author{Diego Garcia Lambas\inst{1,2,3}, Frode K. Hansen \inst{4,1}, Facundo Toscano \inst{1}, Heliana E. Luparello\inst{1} \and Ezequiel F. Boero\inst{1,2}}

\institute{Instituto de Astronomía Teórica y Experimental (IATE), CONICET-UNC, Córdoba, Argentina, \email{diego.garcia.lambas@unc.edu.ar}
\and
Observatorio Astronómico de Córdoba (OAC), UNC, Córdoba, Argentina
\and
Comisión Nacional de Actividades Espaciales (CONAE), Córdoba, Argentina
\and
Institute of Theoretical Astrophysics, University of Oslo, PO Box 1029 Blindern, 0315 Oslo, Norway}

   \authorrunning{Garcia Lambas et al.}
   

   \date{\today}

  \abstract
   {The non-Gaussian Cold Spot surrounded by its hot ring is one of the most striking features of the Cosmic Microwave Background (CMB) and has generated significant attention in the literature. It has been speculated that either new physics or the Integrated Sachs-Wolfe (ISW) effect induced by the presence of a cosmic void at relatively high redshift can account for the observations.}
   {Here we investigate if the newly discovered systematic decrease in CMB temperature in the neighbourhood of nearby galaxies may create such a strong temperature depression. In particular, we note that the largest galaxy group complex in the Local Universe, the Eridanus super-group with its neighbouring groups, is in the Cold Spot area. Our goal is to analyse observational galaxy data to characterise the neighbourhood of the Cold Spot, explore the properties of these galaxies and thereby make a prediction of the galaxy induced CMB temperature decrement in this region.}
   {We use the Planck SMICA maps and the publicly available observational galaxy catalogues 2MRS, 6dF and HIPASS  with information on redshifts, positions, magnitudes, and other astrophysical characteristics, as foreground tracers. We apply previously explored mean temperature profile shapes to model the expected temperature decrement from the galaxies in the Cold Spot area. }
{Even after correcting for the mean low temperature of the Cold Spot region, we find that the temperature decrement around galaxies is significantly stronger than the mean decrement in other parts of the sky. We discuss whether this could be attributed to the fact that the Cold Spot area coincides with one of the regions populated by the most HI deficient galaxies. Modeling the foreground temperature profile associated to mainly spiral galaxies in this location, we find a particularly strong temperature decrement due to the presence of the late--type overabundant largest group complex in the nearby universe. A cold spot shape, which to a large degree overlaps with the CMB Cold Spot, is observed.}
{We conclude that the coincidence of the only nearby spiral rich group complex located in the Cold Spot region, and the success of the modelling performed, adds strong evidence to the existence of a local extragalactic foreground which could account for the observed temperature depression, alleviating the tension with an otherwise Gaussian field expected in the CMB.}

   \keywords{Cosmology: Cosmic Microwave Background Radiation --
               Galaxies:  General --
               Methods: data analysis --
               Methods: statistical
            }
   \maketitle
  
%

\section{Introduction}
\label{sec:introduction}

Over the years, various anomalous features in the Cosmic Microwave Background (CMB) have been identified, including deviations from statistical isotropy and/or Gaussianity and other features which are unexpected when compared to simulations based on the best-fit $\Lambda$CDM cosmologies 
\citep[see][and references therein for an extensive review]{schwarz,aluri}. A possible explanation for these anomalies was recently suggested by \citet{hansen} based on the discovery in \citet{luparello}. There a cross-correlation of nearby galaxy positions with Planck and WMAP foreground subtracted CMB maps showed a statistically significant decrement in the mean temperature fluctuations extending to several Mpc around late type galaxies.

Although the cause of this temperature decrement remains yet poorly understood, a phenomenological modelling of its properties lead to scenario in which most of the anomalies can be remarkably accounted for \citep{hansen}.  In that work, the modelled signal around galaxies shows that the cold spot region \citep{Vielva:2003et} is expected to be one of the most dominant cold areas in the sky. Therefore, we explore here a scenario where a concentration of large late-type galaxies in the local Universe provides an explanation for the CMB Cold Spot. 

The Cold Spot was detected by noticing an excess of kurtosis in the distribution of wavelet coefficients at angular scales of about 300 arcmin on the southern hemisphere \citep{Vielva:2003et}. As stressed in \citet{vielva2010}, the morphological properties of the cold 
spot manifest differently depending on whether one look at it in real space or wavelet space. In wavelet space it has a more or less symmetrical shape when elliptical wavelets are used; instead, in real space it looks like a clump of small cold spots, the most prominent one reaching values near $-350 \mu$K and with sizes of about $1^{\circ}$.  In mean it is characterised by a remarkable $150 \mu$K decrement in the CMB average 
temperature covering an area of about $5^{\circ}$ of radius.  Notably, \cite{cruz} demonstrated that this extreme temperature drop can only be 
accounted for in less than $0.2\%$ of $\Lambda$CDM Gaussian simulations.
Even though the significance of the detection has been debated the strong concerns  raised questions about the underlying physical processes responsible for its existence. Several possible explanations have been put forward, among them the existence of unaccounted systematic effect, oversubstraction of known foregrounds and the contribution of secondary anisotropies like the Sundayev-Zeldovich (SZ).

One plausible explanation for the origin of the Cold Spot is linked to the Integrated Sachs-Wolfe (ISW) effect, arising from a large underdense region along the path of CMB photons. Despite previous suggestions, conclusive evidence regarding whether a particularly large void within the field of view of the CMB Cold Spot can satisfactorily explain this anomaly within the framework of the $\Lambda$CDM model remains elusive. For instance, \cite{Mackenzie} and \cite{owusu} raised concerns about the role of such a supervoid in causing the CMB Cold Spot in this region.


\section{Data}
\label{sec:data}

In this work, we use the publicly available Planck CMB data \citep{planck2018},  in particular the SMICA foreground cleaned map \citep{compsep2018} and the corresponding simulations, as well as the main galaxy sample of 2MRS \citep{2mrs}, 6dF \citep{6df} and HIPASS \citep{hipass} to analyse different properties of the galaxies. 
The 2MRS sample contains spectroscopic observations for 11000 galaxies, generating a redshift catalogue that is $97.6\% $ complete to well-defined limits and covers $91\%$ of the sky. The 6dF footprint area covers more than 17000 deg$^2$ in 2 optical band-passes and has more than 400000 redshift determinations up to $z=0.2$. The HIPASS catalogue forms one of the largest uniform catalogues of HI sources, with 4315 sources identified purely by their HI content.
In addition to the redshifts and positions, we use magnitudes and bj-R colours which allow for a comparison with the data regarding morphology and luminosity. \\
We notice the diverse nature of the galaxy catalogues analysed which comprise objects with different average properties. In effect, while the 2MRS galaxies reflect the stellar content through the near infrared emission associated to stars, the HIPASS catalogue is sensitive to the HI content. It is also seen that 6dF galaxies are generally fainter than 2MRS members.
Thus, the three catalogues trace differently nearby structures, and in the context of this work, they may provide different contribution to the foreground effects as analysed in \cite{luparello} (hereafter referred to as L2023).

%

%
%
%
%




\section{Foreground models: The CS region}
\label{sec:fg_models}

In \cite{hansen} (hereafter referred to as H2023), a model map of the L2023 foreground was created by assigning a linear temperature profile with a fixed depth and radius to each large spiral galaxy in the 2MRS catalogue. A relation between physical galaxy size and profile depth as well as a relation between galaxy density and profile depth/radius was assumed based on observed properties of temperature decrements. In this model, the Cold Spot appeared as one of the most prominent features on the sky. \\
Here we aim to make a more detailed model of only the area around the Cold Spot in order to compare to the CMB Cold Spot. We will make a model based on even less assumptions than in H2023 by not applying any relation between galaxy density and shape of the profile. Although this effect is clearly present, it is difficult to model properly. 

\begin{figure}[h!]
\begin{center}
\includegraphics[width=\columnwidth]{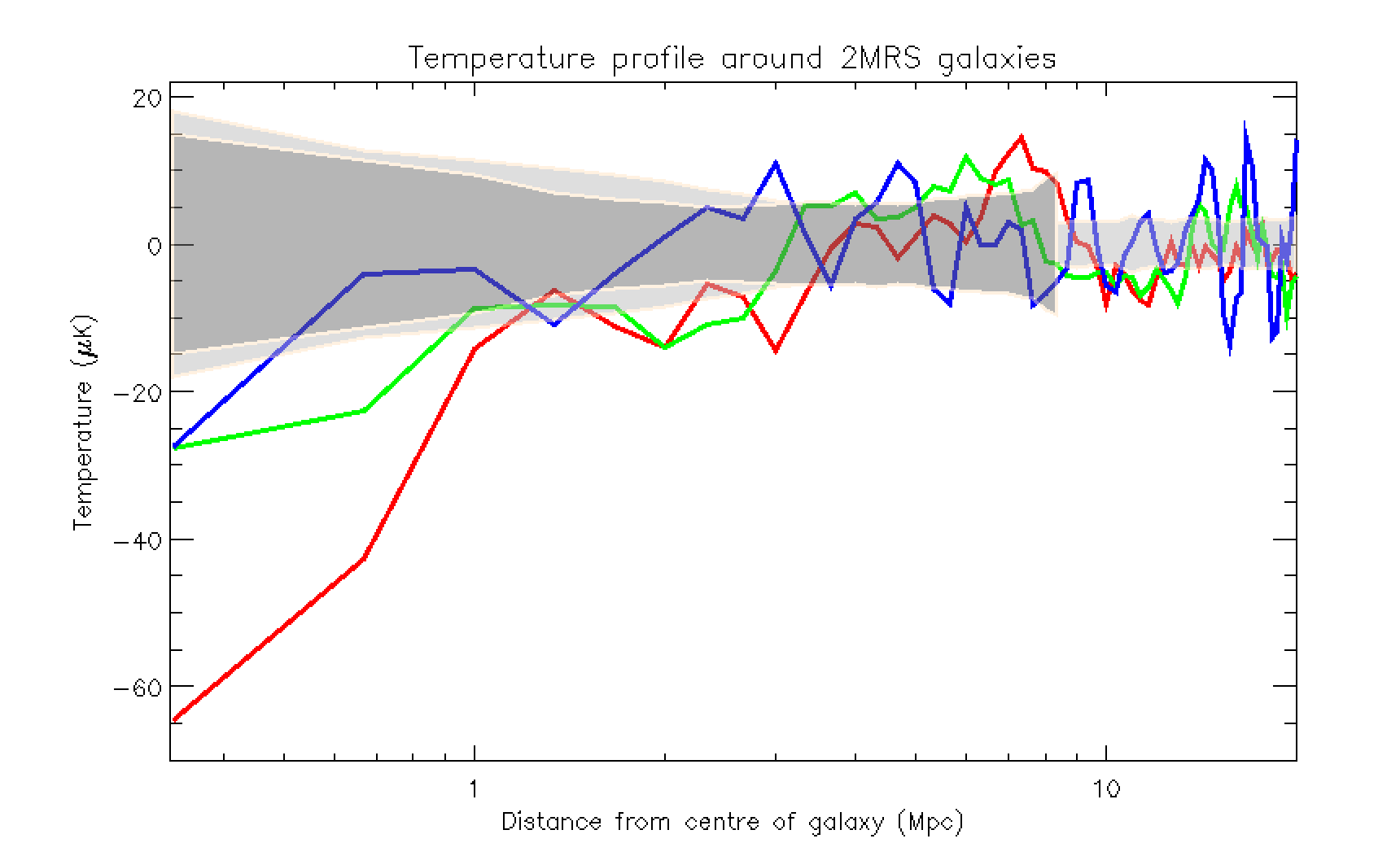}
\end{center}
\caption{\label{fig:profile} Mean temperature profiles around galaxies in the Cold Spot area. Red line: Large ($>8.5$\;kpc) 2MRS spirals within $20^{\circ}$ from the centre of the Cold Spot (12 galaxies). Only galaxies with redshift $z<0.01$ were included. The light grey band shows the $1\sigma$ spread corresponding to 300 simulated CMB maps. The dark gray band shows the profile corresponding to random positions within the Cold Spot area. Green line: HIPASS spirals not including common 2MRS galaxies (22 galaxies). Blue line: 6dF galaxies in the smaller spiral dominated Eridanus and NGC1407 group areas (41 galaxies). The monopole and dipole estimated inside the Cold Spot area was subtracted from the map before these profiles were calculated.}
\end{figure}

In the following, we will do a more detailed modelling of the galaxies in each of the three catalogues. We start by looking at the mean temperature profile of large galaxies within the Cold Spot area defined as a disc with radius $20^{\circ}$ around the Cold Spot. In Figure \ref{fig:profile}, we see the temperature profile of 2MRS galaxies with a depth below $60 \mu$K. This is more than 3 times deeper that the mean depth of profiles taken from the full 2MRS galaxy catalogue in H2023. Notice that the best fit mono- and dipole within the Cold Spot area was subtracted from the map before calculating the profiles such that the mean low temperature of the area is elevated to zero mean temperature. The grey bands show the spread of the profiles taken from 300 simulated maps as well as from 300 random re-positioning of the 2MRS galaxies within the Cold Spot area. Clearly the deep profile of the 2MRS galaxies is at the more than $3\sigma$ level in both cases.

\begin{figure*}[h!]
\begin{center}
\includegraphics[width=0.85\textwidth]{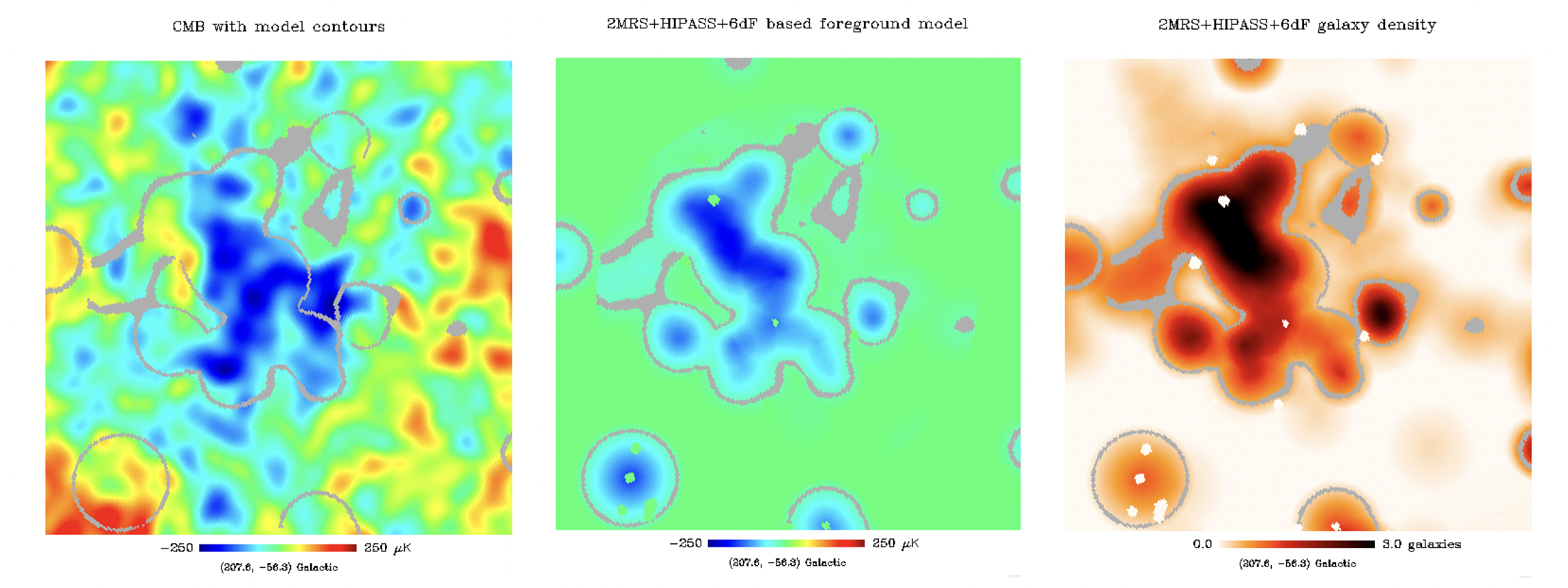}
\end{center}
\caption{\label{fig:model} Predicted foregrounds temperature model in the Cold Spot area. Left panel: CMB Cold Spot area. Middle panel: the temperature model that we obtain using a linear profile around large spiral galaxies in the 2MRS and HIPASS catalogues as well as 6dF galaxies. Right panel: Density of large spiral galaxies where all galaxies in the middle panel have been assigned the same profile depth of 1 (except 6dF galaxies which were assigned a profile depth of 0.2). In the three panels, grey areas indicate $[-20,-30] \mu$K contours to guide the by-eye comparison with the CMB.}
\end{figure*}

The reason for the deeper profiles in this area of the sky is still unknown, but one possible explanation could be found in terms of the degree of HI deficiency as outlined below. However insufficient data about HI deficiency limits the possibility to investigate this further. Therefore, in order to create a realistic model of the Cold Spot, the mean profile we assign to each galaxy cannot be taken from the full sky mean and needs to be calibrated with the actual mean profile within the Cold Spot area. For the three catalogues, we have created model temperature profiles for the galaxies in the following manner:
\begin{itemize}
    \item {\bf 2MRS:} We made a fit of the distribution of all galaxy sizes and profile depths of all spiral galaxies in the 2MRS catalogue outside the Planck common mask area. We found that a power law with index 1.3 made a good fit to the relation between physical galaxy size and profile depth. Using a profile radius of $1.2^{\circ}$ at $z=0.01$, we calibrate the profile depth to the mean profile depth of 2MRS galaxies within the Cold Spot area. As in previous works (L2023, H2023), we found that mainly galaxies with physical size larger than 8.5 kpc contribute to the foreground and we only used these galaxies in the Cold Spot modelling.
    \item {\bf HIPASS:} We first removed all HIPASS galaxies which are also present in the 2MRS catalogue as well as galaxies which are not spiral galaxies. As the HIPASS catalogue does not include physical galaxy sizes, we have used Bj magnitudes for the size-depth relation. We found that only galaxies with Bj magnitude smaller than $-18$ show a clear temperature profile and only these are included in the modelling. In the same manner as for 2MRS we found a relation between Bj magnitudes and profile depths for the full catalogue and we calibrated the profile depth using mean depth in the Cold Spot area (see Figure \ref{fig:profile}).
    \item {\bf 6dF:} We first removed all 6dF galaxies which are also present in the 2MRS and HIPASS. As 6dF has no information on morphology and size, we are unable to make a detailed modelling of large spiral galaxies. However, we know that the Eridanus group as well as the neighbouring NGC1407 group (described below) are dominated by spiral galaxies. For this reason, we obtain the profile of 6dF galaxies in this area (see Figure \ref{fig:profile}) and adjust the profile depth accordingly. We then adopt this profile to all 6dF galaxies in the Cold Spot area, independently of size, magnitude or morphology.
\end{itemize}

Figure \ref{fig:model} shows the predicted temperature decrement in the Cold Spot area using this simplified model based on large 2MRS, HIPASS and 6dF galaxies compared to the CMB Cold Spot area. We also show a similar map of galaxy density where each galaxy was assigned a profile depth of 1 (except 6dF galaxies which were assigned a profile depth of 0.2 as the density of the small 6dF galaxies is very high and dominating). Both in the Cold Spot model and in the galaxy density map, we can clearly see a strong resemblance with the Cold Spot, even in the substructure. The area occupied by the galaxies in this area overlaps to a large degree by the area covered by the CMB Cold Spot. Our model does not take into account the unknown relation between profile depth and galaxy environment which we know exists (H2023). It also uses a very uncertain relation between size/absolute magnitude and profile depth. The 6dF galaxy profiles are even more uncertain, not knowing which galaxies are spiral galaxies. Still, the simplified foreground model gives a predicted Cold Spot structure with a shape and position which coincides remarkably well with the CMB cold spot. \\

\begin{figure*}[h!]
\begin{center}
\includegraphics[width=0.7\textwidth]{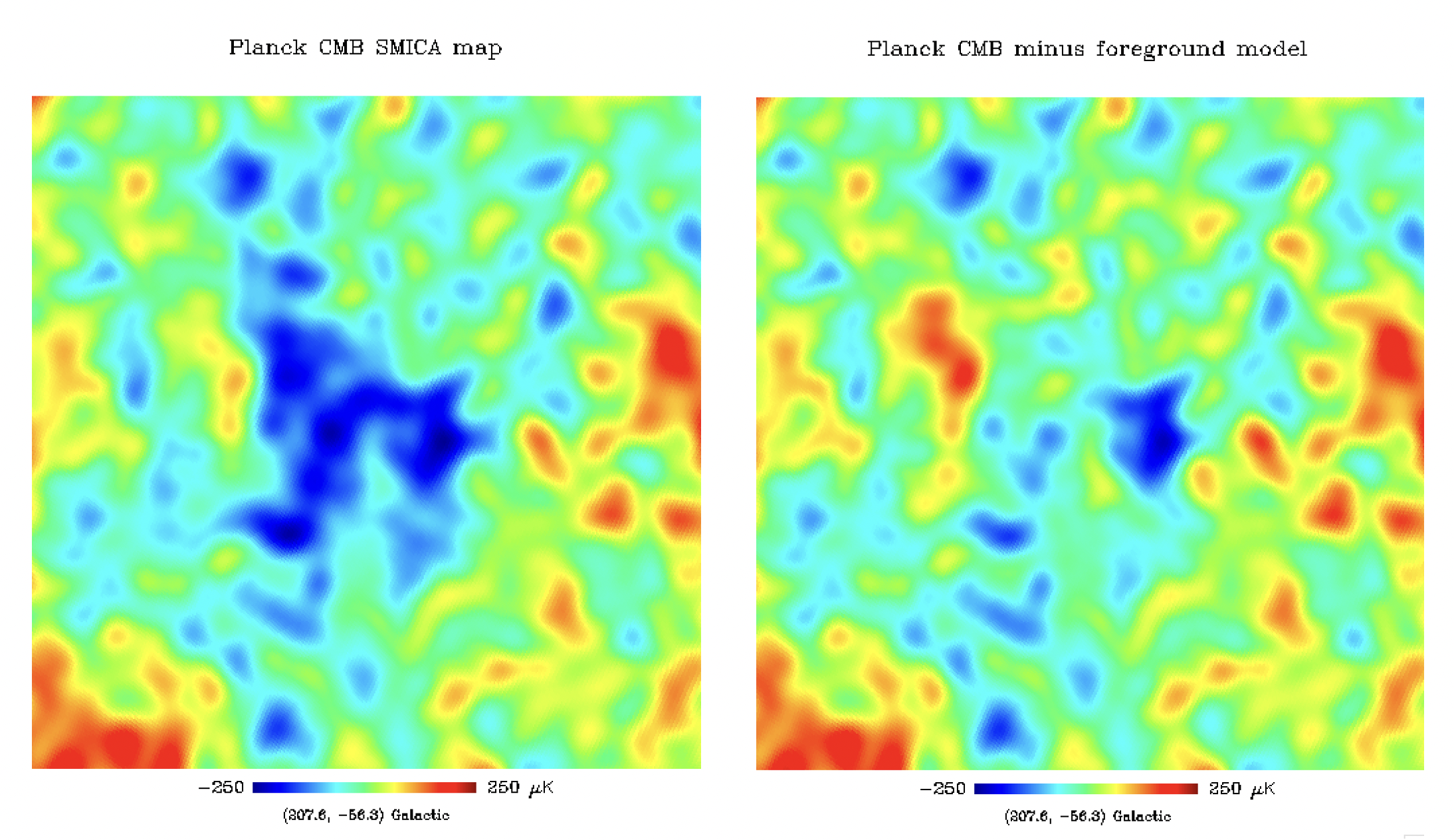}
\end{center}
\caption{\label{fig:subtracted} The Cold Spot area in the Planck SMICA map. It can be seen the map before (left panel) and after (right panel) subtraction of the foregrounds galaxy based model shown in Figure \ref{fig:model}.}
\end{figure*}

In Figure \ref{fig:subtracted}, we show the Cold Spot area before and after subtraction of the model in Figure \ref{fig:model}. We can see that even with this simple model, most of the cold spot signal is removed after this correction. This will be discussed further in the next section.

\begin{figure}[h!]
\begin{center}
\includegraphics[width=\columnwidth]{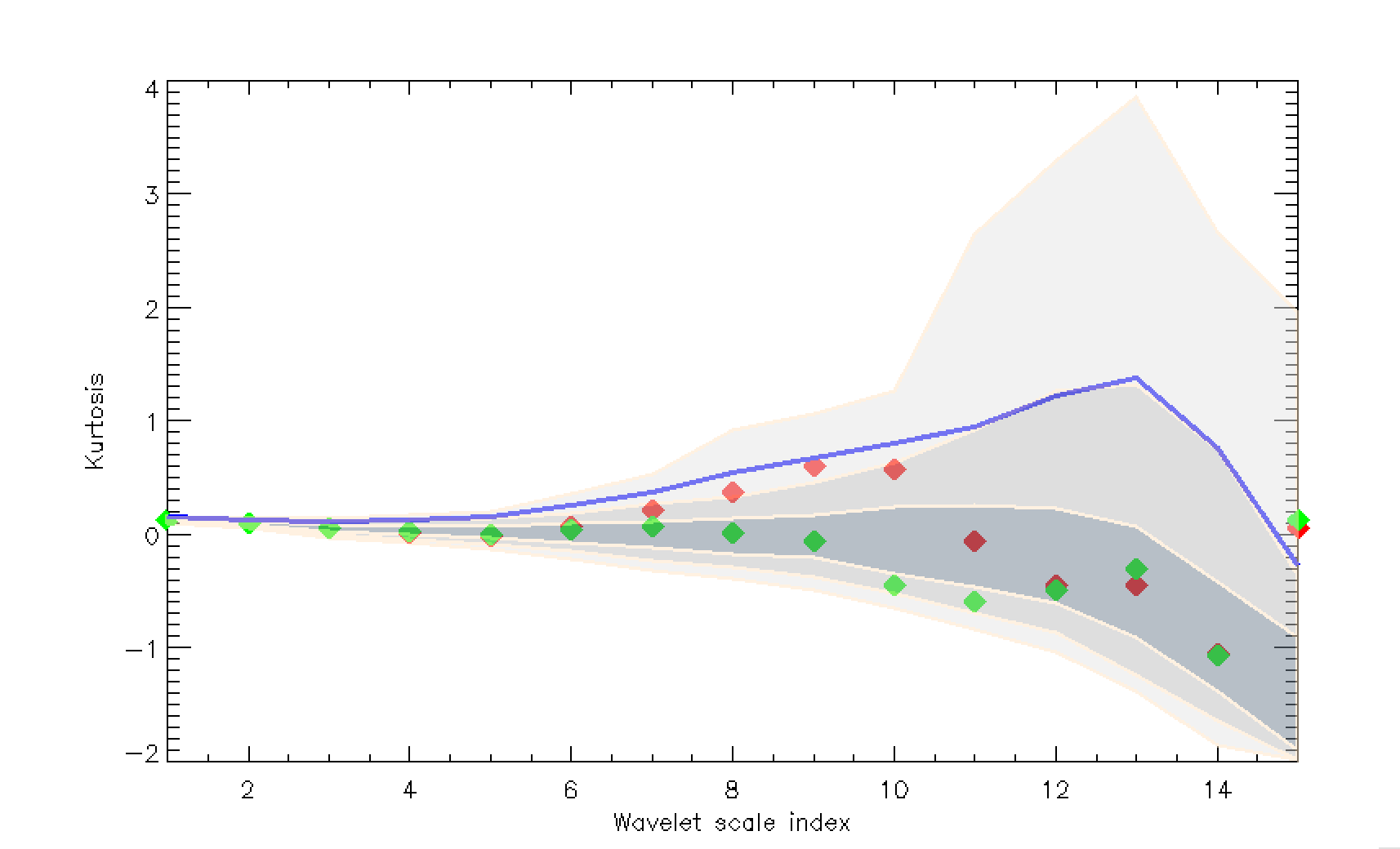}
\end{center}
\caption{\label{fig:kurtosis} Kurtosis of SMHW wavelet coefficients. The grey bands show the 1, 2 and 3$\sigma$ spread of kurtosis in 1000 Gaussian simulations. The red dots show the kurtosis of wavelet coefficients of the Planck SMICA map before correction for the galaxy based model. The green dots show the kurtosis after the correction. The blue line shows the upper limit of the 2$\sigma$ band for Gaussian simulations where the galaxy model was added to each simulation.}
\end{figure}

In Figure \ref{fig:kurtosis}, we show the kurtosis of wavelet coefficients before and after subtraction of the model. We use the same wavelet scales and scale index numbers as in \cite{Vielva:2003et} and also extend the Planck common mask depending on the wavelet scale in a similar manner. Before the subtraction, we can clearly see the original non-Gaussian detection by \cite{Vielva:2003et} around the $2-3\sigma$ level for wavelet scale indices 7, 8, 9 and 10. After the subtraction, the kurtosis for these scales are well within $2\sigma$ while other scales remain practically unchanged. In order to check if our model may induce a kurtosis excess, we also added it to 1000 Gaussian simulations. The result was an increased kurtosis in these simulations for exactly the scale indices 7, 8, 9 and 10. The $2\sigma$ upper spread of the simulations with the model added is shown by the blue line in Figure \ref{fig:kurtosis}. The original $2-3\sigma$ outliers are now well within the $2\sigma$ bands, showing that such high values for the kurtosis is common in the simulations where the galaxy model is added.

\begin{figure}[h!]
\begin{center}
\includegraphics[width=\columnwidth]{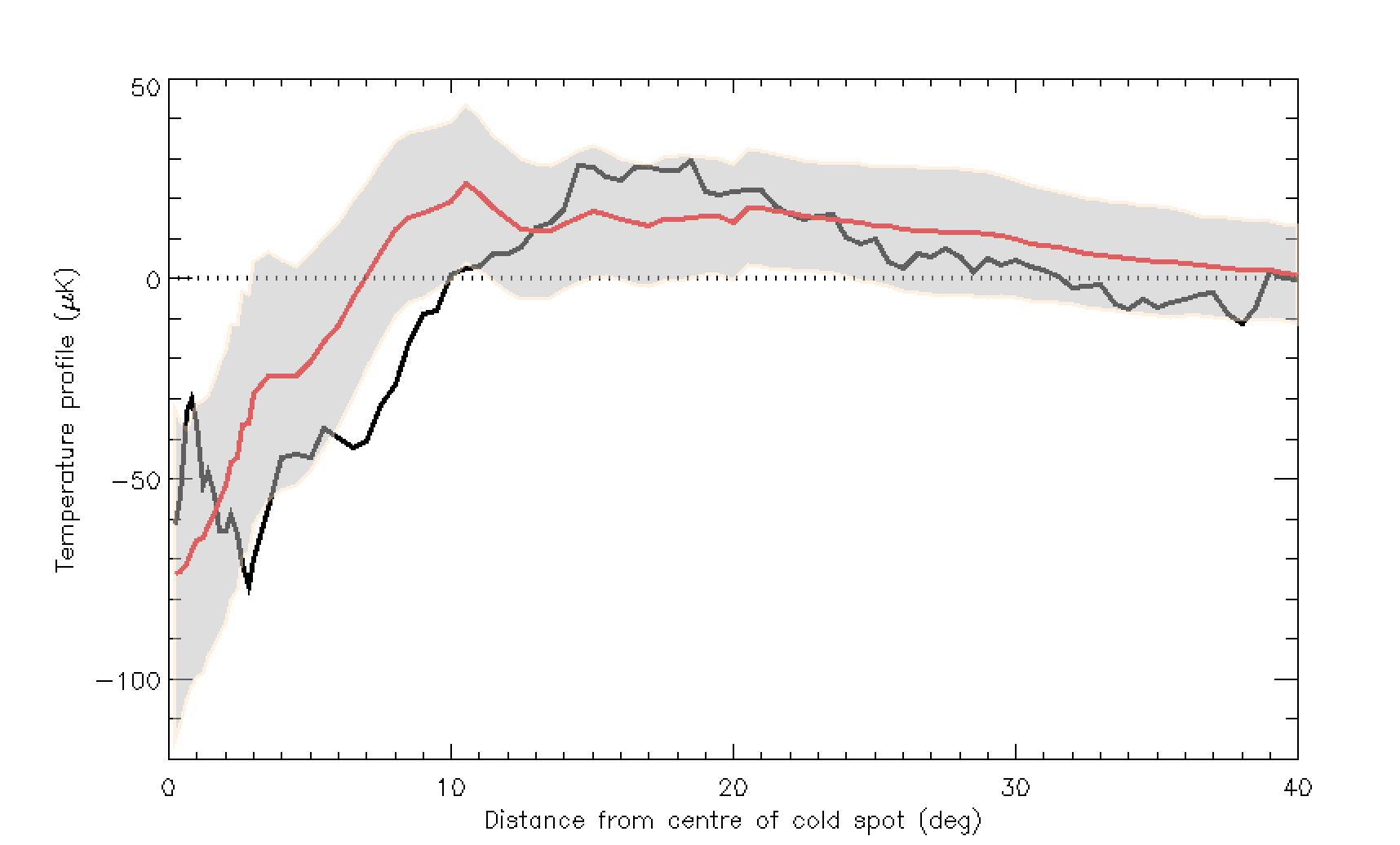}
\end{center}
\caption{\label{fig:hotring} Temperature profile around the centre of the Cold Spot. Black line: Profile from Smica Planck map. Red line: model profile from simulated maps.The grey band shows the standard deviation of the profile from simulations.} 
\end{figure}

Several authors \citep{zhang,cai,Inoue} have noted the presence of a hot ring around the Cold Spot. It has even been claimed that the hot ring is more anomalous than the Cold Spot itself. In Figure \ref{fig:hotring}, the black line shows the temperature profile around the centre of the Cold Spot for the CMB data where we can clearly see the Cold Spot depression and the corresponding hot ring outside. In the same figure, we show the corresponding mean Cold Spot temperature profile for 300 simulated maps where our Cold Spot model has been added to each simulation. Firstly, we note that the angular extension of the Cold Spot in our simplified model is slightly smaller, as can also be seen in Figure \ref{fig:model}. Secondly, a similar hot ring profile can be seen for the simulations outside the Cold Spot depression, but slightly shifted inwards due to the smaller extension of the Cold Spot area. Notice that the mean has been taken only over the half of the simulations where the mean temperature within a $10^{\circ}$ radius is positive. We can therefore see that the model Cold Spot imprinted on a positive CMB fluctuation creates a hot ring around the Cold Spot for half of the simulations.

\section{Nearby galaxies within $z<0.01$, a concentration of late-type galaxies in the Cold Spot region}
\label{sec:galaxies}

\begin{figure}[h!]
\begin{center}
\includegraphics[width=\columnwidth]{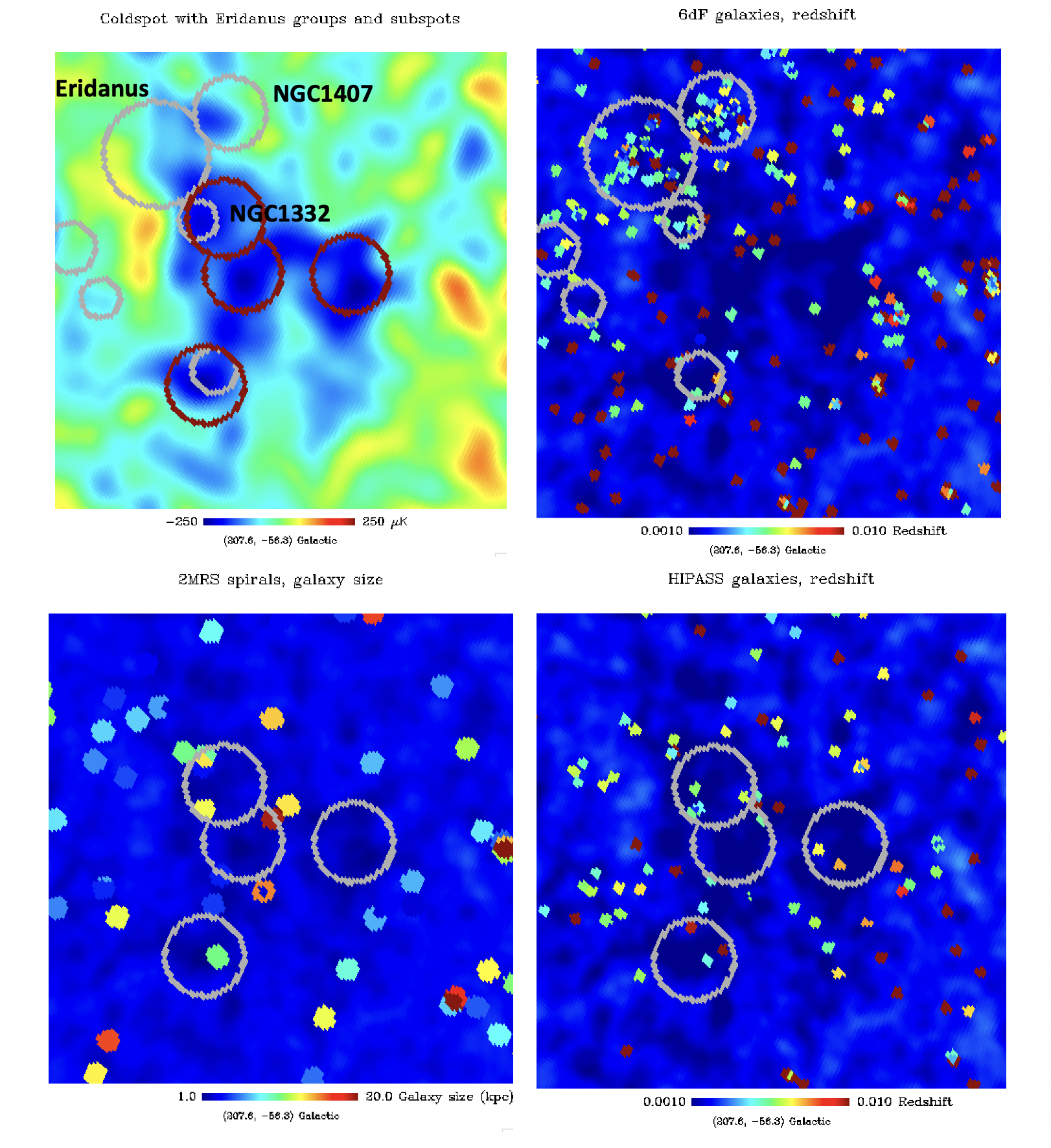}
\end{center}
\caption{\label{fig:CS_gal_groups} Galaxy groups and analysed galaxies in the Cold Spot area. In the upper plots, grey circles show the galaxy groups from Fig. 5 of \cite{brough}. In the lower plots, grey circles show the subspots as defined in \cite{cruz} (in the upper left plot these subspots are shown as red circles).  Upper left panel: the CMB cold spot. Right panels: galaxies 6dF (upper) and HIPASS (lower), galaxies colour coded with redshift. Lower left panel: 2MRS spirals colour coded with galaxy size.}
\end{figure}

\begin{figure}[h!]
\begin{center}
\includegraphics[width=0.9\columnwidth]{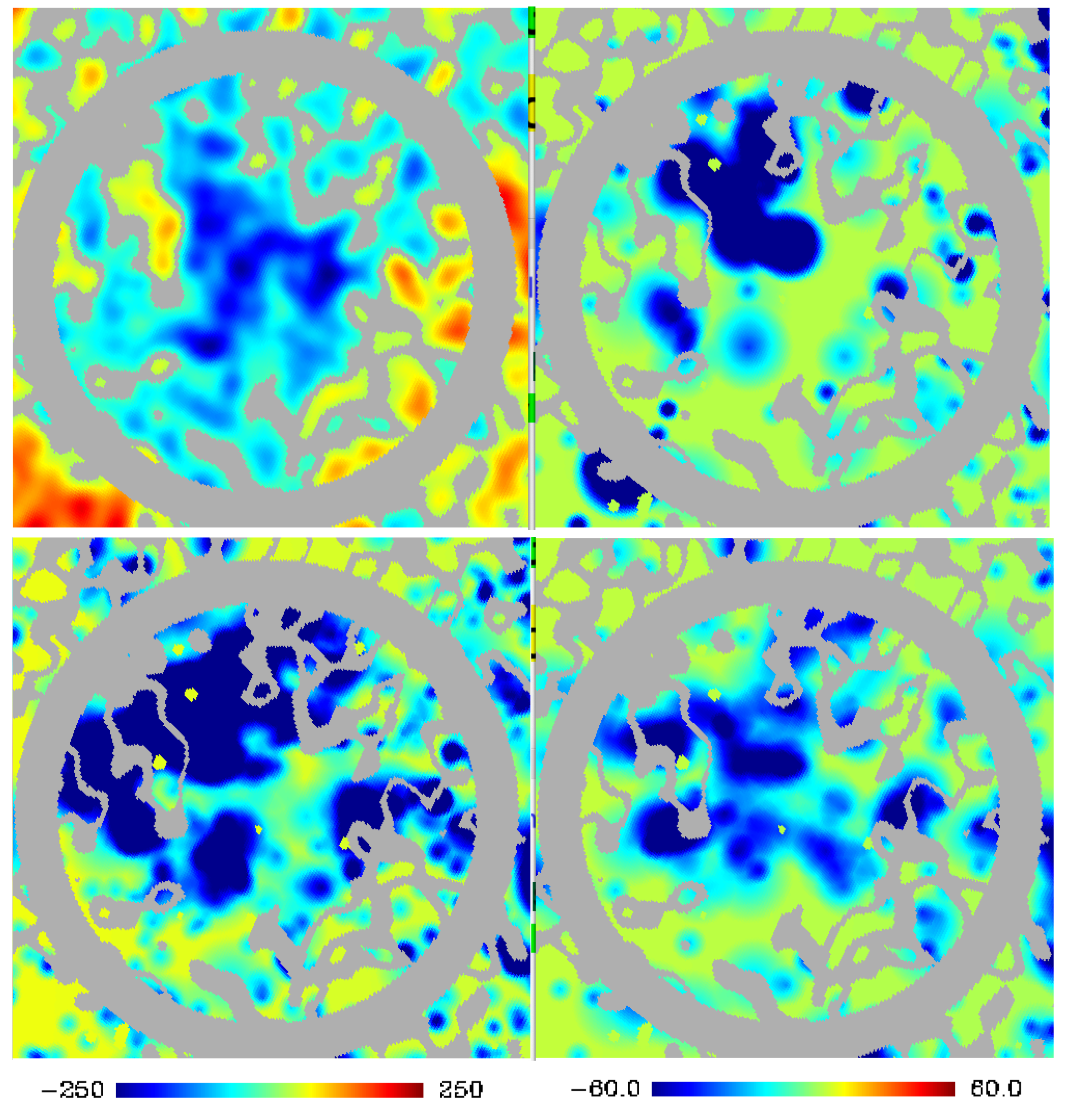}
\end{center}
\caption{\label{fig:foregrounds_CS} Temperature angular distribution in the Cold Spot area. In clockwise direction: CMB Temperature map from Planck 18, 2MRS, HIPASS and 6dF foregrounds. Note that in this figure all galaxies are included and given the same profile as in L2023.}
\end{figure}

It is remarkable that in the nearby Universe, the Cold Spot region is located within the surroundings of the most prominent local galaxy group complex \citep{brough}. This association is  made up of three individual groups in the process of merging onto a future cluster of  mass $\sim7\times10^{13}\mathrm{M}_\odot$. The three groups, NGC 1407, NGC 1332 and the Eridanus group have mutual separations of a few Mpc making them a large overdensity of late-type galaxies.
This group system has a very similar mass to that of the close Fornax cluster. Thus, within the local 50 Mpc$h^{-1}$ structures, this complex is only surpassed in mass by the Virgo cluster.  \\
Having used the galaxies in the Cold Spot area to create a foreground model, we will now go on to identify which galaxies  and groups in the different catalogues could be responsible for the signal. In Figure \ref{fig:CS_gal_groups}, we show the Eridanus complex with NGC1332 and NGC1407 as well as other groups identified in \cite{brough} on top of the CMB temperature field in the Cold Spot area. This plot also shows the 4 coldest regions or subspots as defined in \cite{cruz}. We can see the agreement between some groups of galaxies and subspots of the Cold Spot. Furthermore, in the other panels we show the observed galaxies from the 3 different catalogues used in our analysis. Both 6dF panel and HIPASS panel have a colour code defined by the redshift of the galaxies. In the 6dF panel, we clearly see galaxies clustered around several of these groups with a predominance of $\sim 0.005$ redshifts. On the other hand, the 2MRS catalogue only contain the larger galaxies and only the late type spiral galaxies are shown as these are the ones believed to create the foreground signal (see L2023). In this plot, the colour code is defined by the galaxy size. \\
As the angular extension of the temperature decrement depends upon redshift and the Cold Spot contains several cold substructures extending over several degrees on the sky, we limit the colour scale at $z<0.01$ where the galaxies will produce the most extended structures. The galaxies shown as dark red are therefore beyond this redshift. These still seem to create a foreground signal, but with a smaller angular extension than the nearby galaxies. \\
The region known as the Eridanus complex corresponds to an extended complex of galaxies and galaxy systems whose dynamics and system composition has been a subject of debate 
\citep[see for instance][and references therein]{brough}. This situation arises due to the particular dynamical stage of the cloud of galaxies, probably in the process of accretion onto a future massive cluster. Among the three main groups in the Eridanus complex, only NGC1332 can be associated with one of the 4 strongest temperature decrements of the cold spot (as can be seen in Figure \ref{fig:CS_gal_groups}). However as the figure shows, both the Eridanus group and NGC1407 have visible temperature decrements and form part of the Cold Spot structure.\\
While the upper part of the cold spot seems to be explained by these groups, the lower part only contains one grey circle indicating a small group in this area. While this group coincides with another of the 4 colder areas of the Cold Spot, so does a nearby galaxy which we can see in the 2MRS catalogue. Whether this local temperature depression is resulting from the whole group or mainly from the large galaxy is not clear. The 2MRS catalogue also show several large and nearby spiral galaxies between this group and Eridanus (Figure \ref{fig:CS_gal_groups}). We know that mostly the larger galaxies produce a temperature decrement. \\
%
In 2MRS we can see larger galaxies inside or close to 3 of the subspots. The subspot on the right hand side however, does not have a clear galaxy candidate. However looking at 6dF and HIPASS in Figure \ref{fig:CS_gal_groups}, we can see several galaxies nearby, although their size and properties are unknown. This cold subspot could originate from one or a superposition of several of these, but could also be a cold spot of the CMB itself. In our simplified model (Figure \ref{fig:model}), this subspot is not well reproduced. \\
In order to represent galaxy density from these three catalogues in the Cold Spot area, in Figure \ref{fig:foregrounds_CS}, we have applied the model of L2023 to each of the three catalogues separately. Here we do not use any relation between galaxy size, redshift and shape of the profile. All galaxies are assigned the same profile and in this way, the figure shows a measure of galaxy density, independently of galaxy properties. Note that this figure also include all the smaller galaxies not shown in the previous figures. We confirm again that the upper part of the Cold Spot is dominated by 6dF galaxies with some 2MRS galaxies while in the rightmost part of the Cold Spot there are more HIPASS galaxies.

\section{HI deficiency parameter of foreground galaxies}
\label{sec:deficiency}

\begin{figure}[h!]
\begin{center}
\includegraphics[width=0.6\columnwidth]{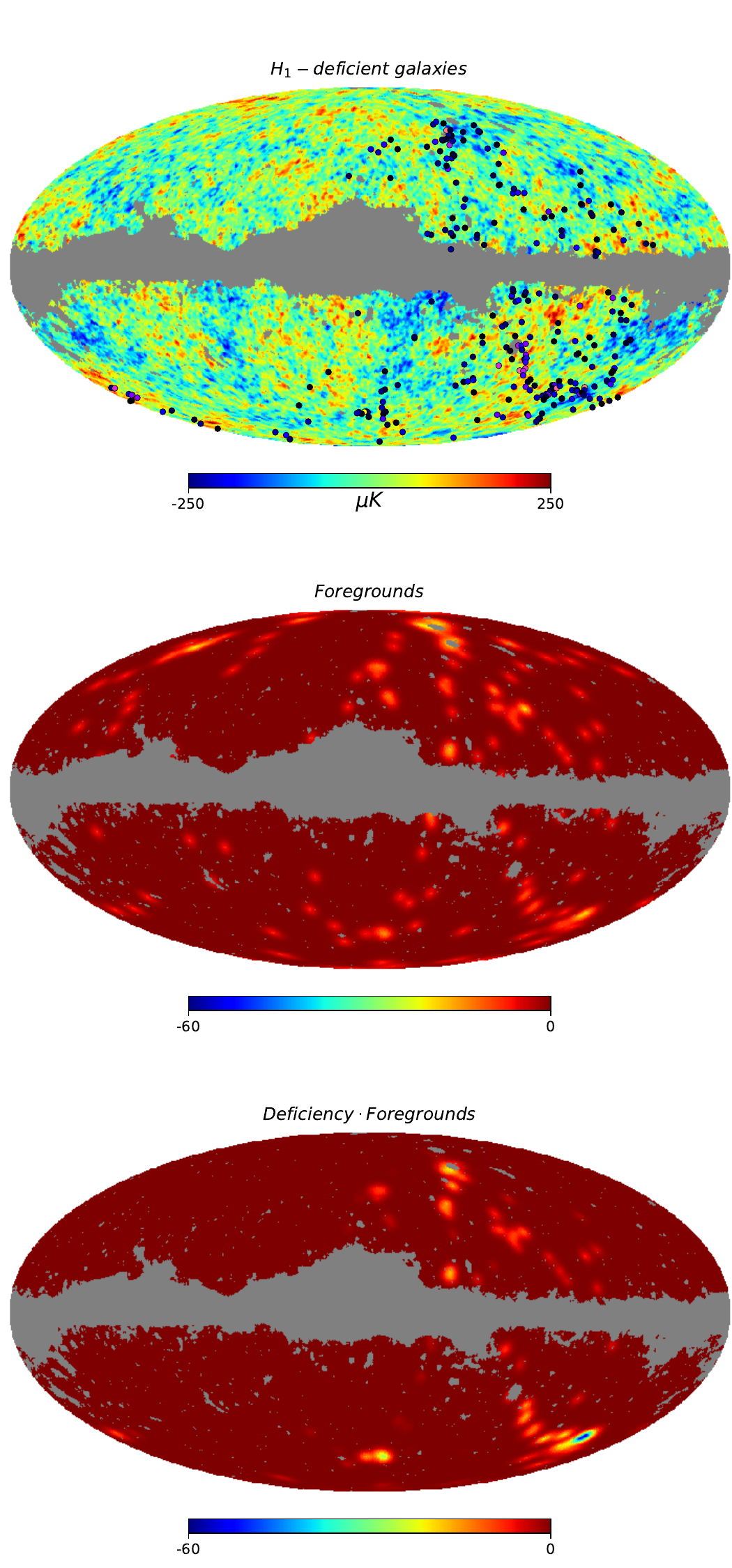}
\end{center}
\caption{\label{fig:CMB_deficient_maps}Foregrounds map weighted by HI deficient galaxies. Upper panel: HI deficient galaxies superposed to the CMB SMICA Temperature map. Middle panel: L2023 Foreground map. Lower panel: L2023 Foreground map weighted by the highest HI deficiency quartile.}
\end{figure}

In a recent study, \cite{2021MNRAS.507.2300F} have analysed the HI content of galaxies belonging to the Eridanus group complex. Among the parameters that are  relevant for our analysis is the Hydrogen deficiency parameter defined as: $DEF_{H_I} = log[M_{H_Iexp}] - log[M_{H_Iobs}]$ providing a measure of the fraction of neutral Hydrogen striped from the galaxies. A visual inspection to the distribution of the HI R-band deficient galaxies in the plane of the sky (see Figure \ref{fig:CMB_deficient_maps}) suggests that the 
Cold Spot region is overabundant in highly deficient HI galaxies. \\ 
We follow \cite{2014MNRAS.444..667D} and calculate the R-band deficiency parameter $DEF_{H_I}$ for galaxies in the redshift range $[0.001, 0.01]$. Then, we consider 4 quartiles in $DEF_{H_I}$ of HIPASS galaxies and construct 4 deficiency weighted maps and  multiply them by L2023 foreground map. We find that the HI deficiency weighting has no significant effect in the original foreground map for the first three quartiles. However, as shown in Figure \ref{fig:CMB_deficient_maps}, the highest quartile makes significantly more prominent the Cold Spot region, giving a hint that stripping of material associated to the Eridanus super group region may be a key ingredient in the generation of the local foreground associated to the Cold Spot.


\section{Summary and conclusions}
\label{sec:conclusions}

We have modelled the CMB temperature decrement in the Cold Spot area assuming a decrease in CMB temperature around large spiral galaxies as in L2023. We find that the galaxy density in the area is particularly high due to the large and nearby Eridanus galaxy group complex. The model predicts a CMB temperature decrement which to a surprisingly large degree resembles the CMB cold spot, both in shape and actual temperatures. \\
We observe that this temperature decrement is about 3 times stronger around individual galaxies in the Cold Spot area than in other locations with the same area in the CMB (after having subtracted the mean temperature around the Cold Spot). Besides, we find that when weighting the full-sky foreground model L2023 by the galaxy density of HI deficient galaxies, the cold spot signal increases by a factor three, indicating that stripping by tidal interactions may be an efficient mechanism in the foreground generation and could explain the deeper temperature profiles around galaxies in this area. \\
The fact that the Cold Spot can be reproduced remarkably well by modelling of the galaxy induced temperature decrement and that most large scale anomalies of the CMB arise naturally when such a model is applied to the full sky (H2023), adds to the increasing evidence for the existence of an unknown CMB foreground component and warrants further investigations of its properties and possible contamination of the CMB.

\begin{acknowledgements}
 This work was partially supported by Agencia Nacional de Promoción Científica y Tecnológica (PICT 2015-3098, PICT 2016-1975), the Consejo Nacional de Investigaciones Científicas y Técnicas (CONICET, Argentina) and the Secretaría de Ciencia y Tecnología de la Universidad Nacional de Córdoba (SeCyT-UNC, Argentina). Results in this paper are based on observations obtained with Planck (http://www.esa.int/Planck), an ESA science mission with instruments and contributions directly funded by ESA Member States, NASA, and Canada. The simulations were performed on resources provided by UNINETT Sigma2 - the National Infrastructure for High Performance Computing and Data Storage in Norway". Some of the results in this paper have been derived using the HEALPix package. \citep{healpix}

\end{acknowledgements}

\bibliographystyle{aa}
\bibliography{references}

\begin{thebibliography}{21}
\expandafter\ifx\csname natexlab\endcsname\relax\def\natexlab#1{#1}\fi

\bibitem[{{Aluri} {et~al.}(2017){Aluri}, {Ralston}, \& {Weltman}}]{aluri}
{Aluri}, P.~K., {Ralston}, J.~P., \& {Weltman}, A. 2017, \mnras, 472, 2410

\bibitem[{{Brough} {et~al.}(2006){Brough}, {Forbes}, {Kilborn}, {Couch}, \&
  {Colless}}]{brough}
{Brough}, S., {Forbes}, D.~A., {Kilborn}, V.~A., {Couch}, W., \& {Colless}, M.
  2006, \mnras, 369, 1351–1374

\bibitem[{{Cai} {et~al.}(2010){Cai}, {Cole}, {Jenkins}, \& {Frenk}}]{cai}
{Cai}, Y.-C., {Cole}, S., {Jenkins}, A., \& {Frenk}, C.~S. 2010, \mnras, 407,
  201

\bibitem[{{Cruz} {et~al.}(2005){Cruz}, {Mart\'{i}nez–Gonz\'alez}, {Vielva},
  \& {Cay\'on}}]{cruz}
{Cruz}, M., {Mart\'{i}nez–Gonz\'alez}, E., {Vielva}, P., \& {Cay\'on}, L.
  2005, \mnras, 356, 29

\bibitem[{{D{\'e}nes} {et~al.}(2014){D{\'e}nes}, {Kilborn}, \&
  {Koribalski}}]{2014MNRAS.444..667D}
{D{\'e}nes}, H., {Kilborn}, V.~A., \& {Koribalski}, B.~S. 2014, \mnras, 444,
  667

\bibitem[{{For} {et~al.}(2021){For}, {Wang}, {Westmeier}, {Wong}, {Murugeshan},
  {Staveley-Smith}, {Courtois}, {Pomar{\`e}de}, {Spekkens}, {Catinella},
  {McQuinn}, {Elagali}, {Koribalski}, {Lee-Waddell}, {Madrid}, {Popping},
  {Reynolds}, {Rhee}, {Bekki}, {D{\`e}nes}, {Kamphuis}, \&
  {Verdes-Montenegro}}]{2021MNRAS.507.2300F}
{For}, B.~Q., {Wang}, J., {Westmeier}, T., {et~al.} 2021, \mnras, 507, 2300

\bibitem[{{G{\'o}rski} {et~al.}(2005){G{\'o}rski}, {Hivon}, {Banday},
  {Wandelt}, {Hansen}, {Reinecke}, \& {Bartelmann}}]{healpix}
{G{\'o}rski}, K.~M., {Hivon}, E., {Banday}, A.~J., {et~al.} 2005, \apj, 622,
  759

\bibitem[{{Hansen} {et~al.}(2023){Hansen}, {Boero}, {Luparello}, \& {Garcia
  Lambas}}]{hansen}
{Hansen}, F.~K., {Boero}, E.~F., {Luparello}, H.~E., \& {Garcia Lambas}, D.
  2023, \aap, 675, L7

\bibitem[{{Huchra} {et~al.}(2012){Huchra}, {Macri}, {Masters}, {Jarrett},
  {Berlind}, {Calkins}, {Crook}, {Cutri}, {Erdo{\v{g}}du}, {Falco}, {George},
  {Hutcheson}, {Lahav}, {Mader}, {Mink}, {Martimbeau}, {Schneider},
  {Skrutskie}, {Tokarz}, \& {Westover}}]{2mrs}
{Huchra}, J.~P., {Macri}, L.~M., {Masters}, K.~L., {et~al.} 2012, \apjs, 199,
  26

\bibitem[{Inoue {et~al.}(2010)Inoue, Sakai, \& Tomita}]{Inoue}
Inoue, K.~T., Sakai, N., \& Tomita, K. 2010, The Astrophysical Journal, 724, 12

\bibitem[{{Jones} {et~al.}(2009){Jones}, {Read}, {Saunders}, {Colless},
  {Jarrett}, {Parker}, {Fairall}, {Mauch}, {Sadler}, {Watson}, {Burton},
  {Campbell}, {Cass}, {Croom}, {Dawe}, {Fiegert}, {Frankcombe}, {Hartley},
  {Huchra}, {James}, {Kirby}, {Lahav}, {Lucey}, {Mamon}, {Moore}, {Peterson},
  {Prior}, {Proust}, {Russell}, {Safouris}, {Wakamatsu}, {Westra}, \&
  {Williams}}]{6df}
{Jones}, D.~H., {Read}, M.~A., {Saunders}, W., {et~al.} 2009, \mnras, 399, 683

\bibitem[{{Luparello} {et~al.}(2023){Luparello}, {Boero}, {Lares},
  {S{\'a}nchez}, \& {Garcia Lambas}}]{luparello}
{Luparello}, H.~E., {Boero}, E.~F., {Lares}, M., {S{\'a}nchez}, A.~G., \&
  {Garcia Lambas}, D. 2023, \mnras, 518, 5643

\bibitem[{{Mackenzie} {et~al.}(2017){Mackenzie}, {Shanks}, {Bremer}, {Cai},
  {Gunawardhana}, {Kov{\'a}cs}, {Norberg}, \& {Szapudi}}]{Mackenzie}
{Mackenzie}, R., {Shanks}, T., {Bremer}, M.~N., {et~al.} 2017, \mnras, 470,
  2328

\bibitem[{{Meyer} {et~al.}(2004){Meyer}, {Zwaan}, {Webster}, {Staveley-Smith},
  {Ryan-Weber}, {Drinkwater}, {Barnes}, {Howlett}, {Kilborn}, {Stevens},
  {Waugh}, {Pierce}, {Bhathal}, {de Blok}, {Disney}, {Ekers}, {Freeman},
  {Garcia}, {Gibson}, {Harnett}, {Henning}, {Jerjen}, {Kesteven}, {Knezek},
  {Koribalski}, {Mader}, {Marquarding}, {Minchin}, {O'Brien}, {Oosterloo},
  {Price}, {Putman}, {Ryder}, {Sadler}, {Stewart}, {Stootman}, \&
  {Wright}}]{hipass}
{Meyer}, M.~J., {Zwaan}, M.~A., {Webster}, R.~L., {et~al.} 2004, \mnras, 350,
  1195

\bibitem[{{Owusu} {et~al.}(2023){Owusu}, {da Silveira Ferreira}, {Notari}, \&
  {Quartin}}]{owusu}
{Owusu}, S., {da Silveira Ferreira}, P., {Notari}, A., \& {Quartin}, M. 2023,
  \jcap, 2023, 040

\bibitem[{{Planck Collaboration} {et~al.}(2020{\natexlab{a}}){Planck
  Collaboration}, {Aghanim}, {Akrami}, {Arroja}, {Ashdown}, {Aumont},
  {Baccigalupi}, {Ballardini}, {Banday}, {Barreiro}, {Bartolo}, {Basak},
  {Battye}, {Benabed}, {Bernard}, {Bersanelli}, {Bielewicz}, {Bock}, {Bond},
  {Borrill}, {Bouchet}, {Boulanger}, {Bucher}, {Burigana}, {Butler},
  {Calabrese}, {Cardoso}, {Carron}, {Casaponsa}, {Challinor}, {Chiang},
  {Colombo}, {Combet}, {Contreras}, {Crill}, {Cuttaia}, {de Bernardis}, {de
  Zotti}, {Delabrouille}, {Delouis}, {D{\'e}sert}, {Di Valentino}, {Dickinson},
  {Diego}, {Donzelli}, {Dor{\'e}}, {Douspis}, {Ducout}, {Dupac}, {Efstathiou},
  {Elsner}, {En{\ss}lin}, {Eriksen}, {Falgarone}, {Fantaye}, {Fergusson},
  {Fernandez-Cobos}, {Finelli}, {Forastieri}, {Frailis}, {Franceschi},
  {Frolov}, {Galeotta}, {Galli}, {Ganga}, {G{\'e}nova-Santos}, {Gerbino},
  {Ghosh}, {Gonz{\'a}lez-Nuevo}, {G{\'o}rski}, {Gratton}, {Gruppuso},
  {Gudmundsson}, {Hamann}, {Handley}, {Hansen}, {Helou}, {Herranz},
  {Hildebrandt}, {Hivon}, {Huang}, {Jaffe}, {Jones}, {Karakci}, {Keih{\"a}nen},
  {Keskitalo}, {Kiiveri}, {Kim}, {Kisner}, {Knox}, {Krachmalnicoff}, {Kunz},
  {Kurki-Suonio}, {Lagache}, {Lamarre}, {Langer}, {Lasenby}, {Lattanzi},
  {Lawrence}, {Le Jeune}, {Leahy}, {Lesgourgues}, {Levrier}, {Lewis},
  {Liguori}, {Lilje}, {Lilley}, {Lindholm}, {L{\'o}pez-Caniego}, {Lubin}, {Ma},
  {Mac{\'\i}as-P{\'e}rez}, {Maggio}, {Maino}, {Mandolesi}, {Mangilli},
  {Marcos-Caballero}, {Maris}, {Martin}, {Martinelli},
  {Mart{\'\i}nez-Gonz{\'a}lez}, {Matarrese}, {Mauri}, {McEwen}, {Meerburg},
  {Meinhold}, {Melchiorri}, {Mennella}, {Migliaccio}, {Millea}, {Mitra},
  {Miville-Desch{\^e}nes}, {Molinari}, {Moneti}, {Montier}, {Morgante}, {Moss},
  {Mottet}, {M{\"u}nchmeyer}, {Natoli}, {N{\o}rgaard-Nielsen}, {Oxborrow},
  {Pagano}, {Paoletti}, {Partridge}, {Patanchon}, {Pearson}, {Peel}, {Peiris},
  {Perrotta}, {Pettorino}, {Piacentini}, {Polastri}, {Polenta}, {Puget},
  {Rachen}, {Reinecke}, {Remazeilles}, {Renault}, {Renzi}, {Rocha}, {Rosset},
  {Roudier}, {Rubi{\~n}o-Mart{\'\i}n}, {Ruiz-Granados}, {Salvati}, {Sandri},
  {Savelainen}, {Scott}, {Shellard}, {Shiraishi}, {Sirignano}, {Sirri},
  {Spencer}, {Sunyaev}, {Suur-Uski}, {Tauber}, {Tavagnacco}, {Tenti},
  {Terenzi}, {Toffolatti}, {Tomasi}, {Trombetti}, {Valiviita}, {Van Tent},
  {Vibert}, {Vielva}, {Villa}, {Vittorio}, {Wandelt}, {Wehus}, {White},
  {White}, {Zacchei}, \& {Zonca}}]{planck2018}
{Planck Collaboration}, {Aghanim}, N., {Akrami}, Y., {et~al.}
  2020{\natexlab{a}}, \aap, 641, A1

\bibitem[{{Planck Collaboration} {et~al.}(2020{\natexlab{b}}){Planck
  Collaboration}, {Aghanim}, {Akrami}, {Ashdown}, {Aumont}, {Baccigalupi},
  {Ballardini}, {Banday}, {Barreiro}, {Bartolo}, {Basak}, {Battye}, {Benabed},
  {Bersanelli}, {Bielewicz}, {Bond}, {Borrill}, {Bouchet}, {Boulanger},
  {Bucher}, {Burigana}, {Butler}, {Calabrese}, {Cardoso}, {Carron},
  {Casaponsa}, {Challinor}, {Chiang}, {Colombo}, {Combet}, {Contreras},
  {Crill}, {Cuttaia}, {de Bernardis}, {de Zotti}, {Delabrouille}, {Delouis},
  {D{\'e}sert}, {Di Valentino}, {Dickinson}, {Diego}, {Donzelli}, {Dor{\'e}},
  {Douspis}, {Ducout}, {Dupac}, {Efstathiou}, {Elsner}, {En{\ss}lin},
  {Eriksen}, {Falgarone}, {Fantaye}, {Fergusson}, {Fernandez-Cobos}, {Finelli},
  {Forastieri}, {Frailis}, {Franceschi}, {Frolov}, {Galeotta}, {Galli},
  {Ganga}, {G{\'e}nova-Santos}, {Gerbino}, {Ghosh}, {Gonz{\'a}lez-Nuevo},
  {G{\'o}rski}, {Gratton}, {Gruppuso}, {Gudmundsson}, {Hamann}, {Handley},
  {Hansen}, {Helou}, {Herranz}, {Hildebrandt}, {Hivon}, {Huang}, {Jaffe},
  {Jones}, {Karakci}, {Keih{\"a}nen}, {Keskitalo}, {Kiiveri}, {Kim}, {Kisner},
  {Knox}, {Krachmalnicoff}, {Kunz}, {Kurki-Suonio}, {Lagache}, {Lamarre},
  {Langer}, {Lasenby}, {Lattanzi}, {Lawrence}, {Le Jeune}, {Leahy},
  {Lesgourgues}, {Levrier}, {Lewis}, {Liguori}, {Lilje}, {Lilley}, {Lindholm},
  {L{\'o}pez-Caniego}, {Lubin}, {Ma}, {Mac{\'\i}as-P{\'e}rez}, {Maggio},
  {Maino}, {Mandolesi}, {Mangilli}, {Marcos-Caballero}, {Maris}, {Martin},
  {Martinelli}, {Mart{\'\i}nez-Gonz{\'a}lez}, {Matarrese}, {Mauri}, {McEwen},
  {Meerburg}, {Meinhold}, {Melchiorri}, {Mennella}, {Migliaccio}, {Millea},
  {Mitra}, {Miville-Desch{\^e}nes}, {Molinari}, {Moneti}, {Montier},
  {Morgante}, {Moss}, {Mottet}, {M{\"u}nchmeyer}, {Natoli},
  {N{\o}rgaard-Nielsen}, {Oxborrow}, {Pagano}, {Paoletti}, {Partridge},
  {Patanchon}, {Pearson}, {Peel}, {Peiris}, {Perrotta}, {Pettorino},
  {Piacentini}, {Polastri}, {Polenta}, {Puget}, {Rachen}, {Reinecke},
  {Remazeilles}, {Renault}, {Renzi}, {Rocha}, {Rosset}, {Roudier},
  {Rubi{\~n}o-Mart{\'\i}n}, {Ruiz-Granados}, {Salvati}, {Sandri}, {Savelainen},
  {Scott}, {Shellard}, {Shiraishi}, {Sirignano}, {Sirri}, {Spencer}, {Sunyaev},
  {Suur-Uski}, {Tauber}, {Tavagnacco}, {Tenti}, {Terenzi}, {Toffolatti},
  {Tomasi}, {Trombetti}, {Valiviita}, {Van Tent}, {Vibert}, {Vielva}, {Villa},
  {Vittorio}, {Wandelt}, {Wehus}, {White}, {White}, {Zacchei}, \&
  {Zonca}}]{compsep2018}
{Planck Collaboration}, {Aghanim}, N., {Akrami}, Y., {et~al.}
  2020{\natexlab{b}}, \aap, 641, A4

\bibitem[{{Schwarz} {et~al.}(2016){Schwarz}, {Copi}, {Huterer}, \&
  {Starkman}}]{schwarz}
{Schwarz}, D.~J., {Copi}, C.~J., {Huterer}, D., \& {Starkman}, G.~D. 2016,
  Classical and Quantum Gravity, 33, 184001

\bibitem[{{Vielva}(2010)}]{vielva2010}
{Vielva}, P. 2010, Advances in Astronomy, 2010, 592094

\bibitem[{Vielva {et~al.}(2004)Vielva, Martinez-Gonzalez, Barreiro, Sanz, \&
  Cayon}]{Vielva:2003et}
Vielva, P., Martinez-Gonzalez, E., Barreiro, R.~B., Sanz, J.~L., \& Cayon, L.
  2004, Astrophys. J., 609, 22

\bibitem[{{Zhang} \& {Huterer}(2010)}]{zhang}
{Zhang}, R. \& {Huterer}, D. 2010, Astroparticle Physics, 33, 69

\end{thebibliography}

\end{document}